\documentclass[10pt,twocolumn]{article}
\newif\ifreview \reviewfalse
\usepackage[a4paper,top=2.2cm,bottom=2.4cm,left=1.9cm,right=1.9cm,columnsep=0.65cm,headsep=0.7cm,footskip=1cm]{geometry}
\usepackage[T1]{fontenc}
\IfFileExists{XCharter.sty}{\usepackage{XCharter}}{\usepackage{charter}}
\IfFileExists{sourcesanspro.sty}{\usepackage[scale=0.9]{sourcesanspro}}{\usepackage[scaled=0.88]{helvet}}
\usepackage{microtype}
\usepackage{xcolor}
\usepackage{graphicx}
\usepackage{booktabs}
\usepackage{setspace}
\usepackage[numbers,square,comma,sort&compress]{natbib}
\usepackage{titlesec}
\usepackage{needspace}
\usepackage{fancyhdr}
\usepackage{caption}
\usepackage{lineno}
\usepackage{array}
\usepackage{amsmath,amssymb}
\usepackage{newunicodechar}
\newunicodechar{±}{\ensuremath{\pm}}
\newunicodechar{×}{\ensuremath{\times}}
\newunicodechar{−}{\ensuremath{-}}
\newunicodechar{–}{--}
\newunicodechar{—}{---}
\usepackage[hidelinks,pdftitle={DosePET: An Open-Source Dataset with 2100 Radiation Dose Maps for PET/CT},pdfauthor={Eleftherios Tzanis, Antonios Tzortzakakis, Michail E. Klontzas}]{hyperref}
\ifreview \linenumbers \else  \fi

\titleformat{\section}{\needspace{4\baselineskip}\sffamily\bfseries\small}{}{0pt}{\MakeUppercase}
\titlespacing*{\section}{0pt}{2.4ex plus .5ex minus .2ex}{0.9ex plus .1ex}
\titleformat{\subsection}[runin]{\normalfont\bfseries}{}{0pt}{}[.]
\titlespacing*{\subsection}{0pt}{1.4ex plus .3ex minus .2ex}{0.5em}
\titleformat{\subsubsection}{\needspace{3\baselineskip}\normalfont\bfseries}{}{0pt}{}
\titlespacing*{\subsubsection}{0pt}{1.6ex plus .4ex minus .2ex}{0.6ex plus .1ex}
\DeclareCaptionLabelSeparator{bar}{\enspace|\enspace}
\fancypagestyle{first}{\fancyhf{}\fancyfoot[R]{\sffamily\scriptsize\thepage}}
\newcommand{\FDG}{{[}\textsuperscript{18}F]FDG}
\newcommand{\FPSMA}{{[}\textsuperscript{18}F]PSMA-1007}
\newcommand{\GaPSMA}{{[}\textsuperscript{68}Ga]Ga-PSMA-11}
\urldef{\doseactorurl}\url{https://opengate.readthedocs.io/en/v9.4/tools_to_interact_with_the_simulation_actors.html#dose-measurement-doseactor}
\begin{document}
\thispagestyle{first}
\twocolumn[{%
\begin{flushleft}
{\fontsize{24}{28}\selectfont DosePET: An Open-Source Dataset with 2100 Radiation Dose Maps for PET/CT\par}
\vspace{1.1em}
{\sffamily\large\setstretch{1.25}
Eleftherios~Tzanis\textsuperscript{1,2,*},
Antonios~Tzortzakakis\textsuperscript{1,3,4},
Michail~E.~Klontzas\textsuperscript{1,2}\par}
\end{flushleft}
\vspace{0.6em}
\noindent\rule{\textwidth}{0.6pt}\par\vspace{0.9em}
\noindent\begin{minipage}[t]{0.25\textwidth}{\sffamily\bfseries\small Abstract\par}\end{minipage}%
\begin{minipage}[t]{0.75\textwidth}
\small\setstretch{1.05}\setlength{\parskip}{0.45em}\setlength{\parindent}{0pt}
Large collections of three-dimensional radiation dose maps linked to
patient anatomy are needed to develop artificial intelligence (AI) models
for personalized internal dosimetry in nuclear medicine, but access to
such data remains limited. Here we present DosePET, an open-source
dataset of 2,100 patient-specific absorbed-dose maps generated by Monte
Carlo simulation of publicly available PET/CT examinations: 1,503 \FDG{}
examinations from ENHANCE.PET and 369 \FPSMA{} and 228 \GaPSMA{}
examinations from PSMA-PET-CT-Lesions, acquired at five institutions with
seven scanner models. For each examination, the CT was resampled to the
native PET grid to build a voxelized patient phantom, the quantitative
PET image defined the radionuclide source, and radiation transport was
simulated with GATE version 9.4. Organ and tissue masks were generated
from the CT with TotalSegmentator. Each study folder contains a NIfTI
dose map integrated from scan time to infinity, 117 binary organ/tissue
masks, a report of regional dosimetric quantities, including mean
absorbed doses, dose conversion factors and scan-time dose rates, and
simulation metadata. The total processing time was 1,542.4 hours.
Python tools support selective
downloading and conversion to scan-time dose-rate or dose-per-decay maps,
and an agent skill guides pairing with the source PET/CT images. DosePET
is publicly available at \url{https://huggingface.co/datasets/eltzanis/DosePET}.

{\footnotesize\textbf{Keywords:} Nuclear medicine, Personalized Dosimetry, Monte Carlo Simulation, PET/CT, Medical Imaging Datasets.\par}

{\footnotesize\textbf{Abbreviations:} \textbf{DICOM}: Digital Imaging and Communications in Medicine, \textbf{FDG}: fluorodeoxyglucose, \textbf{GATE}: Geant4 Application for Tomographic Emission, \textbf{GPU}: graphics processing unit, \textbf{HU}: Hounsfield units, \textbf{NIfTI}: Neuroimaging Informatics Technology Initiative, \textbf{PSMA}: prostate-specific membrane antigen.\par}
\end{minipage}
\par\vspace{1.0em}
\noindent\rule{\textwidth}{0.3pt}\par\vspace{1.6em}
}]
{\let\thefootnote\relax
\footnotetext{\noindent\sffamily\scriptsize\setstretch{1.0}%
\textsuperscript{1}Division of Radiology, Department for Clinical Science, Intervention and Technology (CLINTEC), Karolinska Institutet, Stockholm, Sweden.
\textsuperscript{2}Artificial Intelligence and Translational Imaging (ATI) Lab, Department of Radiology, School of Medicine, University of Crete, Heraklion, Crete, Greece.
\textsuperscript{3}Department of Nuclear Medicine and Medical Physics, Section Nuclear Medicine Huddinge, Karolinska University Hospital, Stockholm, Sweden.
\textsuperscript{4}Department of Nuclear Medicine and Medical Physics, Theranostics Trial Center, Karolinska University Hospital, Stockholm, Sweden.
\textsuperscript{*}Corresponding author: Eleftherios Tzanis, PhD, Division of Radiology, Department for Clinical Science, Intervention and Technology (CLINTEC), Karolinska Institutet, Stockholm, Sweden. E-mail: eleftherios.tzanis@ki.se; etzanis@uoc.gr; tzaniseleftherios@gmail.com. ORCID: 0000-0003-0353-481X.}}

\section*{Introduction}

Medical AI model development depends on the availability of large,
well-annotated datasets. Publicly available imaging resources support
tasks such as anatomical segmentation, disease classification and lesion
detection \citep{ref1,ref2}. However, access to large collections of
three-dimensional radiation dose maps linked to patient anatomy remains a
challenge for AI research in personalised nuclear medicine dosimetry
\citep{ref3,ref4,ref5}.

PET/CT examinations expose organs and tissues to considerable radiation
doses \citep{ref6}. Personalised dosimetry can inform radiation
protection and protocol optimisation by accounting for individual anatomy
and activity distributions \citep{ref4}. Deep learning has shown
potential for predicting radiation dose distributions, but published
proof-of-concept studies have used limited patient cohorts
\citep{ref3,ref4,ref7,ref8}. Generating patient-specific Monte Carlo (MC)
radiation dose maps requires image preparation, and radiation transport
simulation, with a substantial computational burden that complicates the
creation of large training datasets \citep{ref3,ref4}. Limited
availability of dose-labelled data remains a challenge for AI-based
dosimetry \citep{ref5}.

In this study, we present DosePET, a dataset of 2,100 three-dimensional
absorbed-dose maps generated through patient-specific Monte Carlo
simulations of publicly available \FDG{}, \FPSMA{} and \GaPSMA{} PET/CT
examinations. The maps describe radiation dose from the PET radionuclides
and are accompanied by anatomical masks, organ/tissue dose reports and
simulation metadata. This resource is intended to support the
development and evaluation of AI models for personalized internal
radiation dosimetry.

\section*{Materials and Methods}

\subsection*{PET/CT datasets}
Two publicly available PET/CT datasets were used for this study:
ENHANCE.PET, comprising 1,683 anonymised whole-body \FDG{} PET/CT
examinations \citep{ref9}, and PSMA-PET-CT-Lesions, comprising 597
whole-body examinations (369 \FPSMA{} and 228 \GaPSMA{}) from 378
patients \citep{ref10}. No ethics committee approval was required for the
secondary analysis of these anonymised data.

The inclusion criteria were the availability of paired PET and CT images,
radionuclide identity, administered activity and quantitative PET voxel
values expressed in BQML (Bq/mL). ENHANCE.PET images were provided as
compressed NIfTI files; acquisition metadata were retrieved from the
accompanying spreadsheets, and image geometry was obtained from the NIfTI
headers. PSMA-PET-CT-Lesions images were provided in DICOM format, and
the required acquisition and spatial metadata were retrieved from the
DICOM headers. After excluding 180 ENHANCE.PET examinations with PET
voxel values expressed in counts (CNTS), 2,100 examinations were included
for further processing: 1,503 \FDG{}, 369 \FPSMA{}, and 228 \GaPSMA{}.
The selected \FDG{} examinations covered malignant and inflammatory
disorders, with additional scans classified as healthy controls or
negative findings in the source metadata \citep{ref9}. The \FPSMA{} and
\GaPSMA{} cohort comprised patients with suspected or diagnosed prostate
cancer who were imaged for initial staging, treatment monitoring or
assessment for recurrence \citep{ref10}. The included examinations
originated from five institutions and seven different PET/CT scanner
models. Participant and examination characteristics are summarised in
\textbf{Table~\ref{tab:characteristics}}.

\subsection*{Monte Carlo simulation}
For each of the 2,100 PET/CT examinations included in this study,
patient-specific MC radiation dosimetry was performed using GATE version
9.4 \citep{ref11}. A Python pipeline automated image preprocessing,
generation of simulation inputs and macros, and simulation execution.
Using the supplied image geometry, CT images were resampled onto the
native PET grid with an identity transform in physical coordinates and
Lanczos windowed-sinc interpolation. CT Hounsfield units (HUs) were
converted to material compositions and mass densities using GATE's HU
material generator and the specified material and density tables to
construct a voxelized patient phantom. PET activity concentrations
(Bq/mL) were multiplied by the voxel volume and converted to MBq per
voxel. The matched CT and activity volumes were exported in MetaImage
format (.mhd/.raw) for simulation. Radionuclide identity and the recorded
physical half-life defined the F-18 or Ga-68 ion source, with source
locations sampled according to the voxel activities. Radiation transport
used the Geant4 emstandard\_opt4 electromagnetic physics list, with
RadioactiveDecay enabled for GenericIon. Absorbed dose was scored on the
PET grid using the DoseActor mass-weighting algorithm and normalized
using the map-total cumulated activity and completed simulation history
count. CT-based organ and tissue segmentation was performed using
TotalSegmentator \citep{ref1}. The resulting binary masks were used to
extract organ- and tissue-specific dosimetric quantities, including mean
absorbed dose, dose conversion factors and scan-time dose rates, from the
generated three-dimensional dose maps. Further details of the simulation
pipeline and dosimetric calculations are provided in the Supplementary
Material. \textbf{Figure~\ref{fig:workflow}} shows the workflow.

Parallel processing was used to accelerate the Monte Carlo simulations,
with 10\textsuperscript{7} primary histories requested per examination.
Processing was performed on a system with an Intel Core i9-14900 CPU,
128 GB RAM and an NVIDIA RTX 5090 GPU.

\section*{Results}

\subsection*{Dataset contents and derived quantities}
Three-dimensional absorbed-dose maps and AI-generated organ/tissue masks
were produced for the 2,100 included PET/CT examinations. The total
processing time across all examinations was 1,542.4 hours (approximately
64 days).

\begin{figure}[p]
\centering
\includegraphics[width=\linewidth,height=0.78\textheight,keepaspectratio]{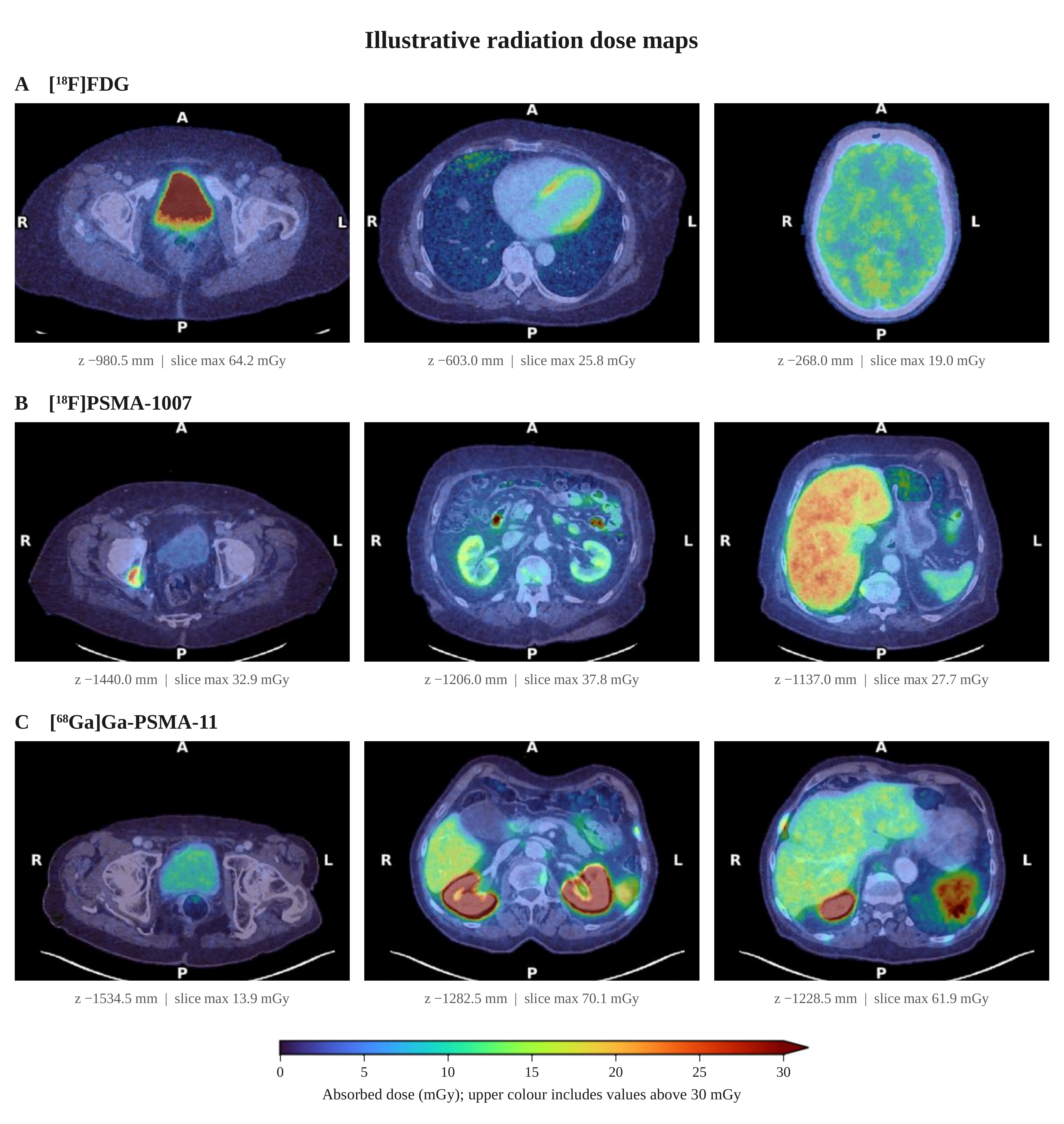}
\caption{Illustrative axial CT-dose overlays for \FDG{} (row A; case FDG/0001), \FPSMA{} (row B; case 40922), and \GaPSMA{} (row C; case 77322). Dose is integrated from scan time to infinity assuming physical decay alone and an unchanged spatial activity distribution. All panels use a common 0-30 mGy color scale, with values above 30 mGy shown in the upper color.}
\label{fig:overlays}
\end{figure}

Each study folder contains a compressed NIfTI dose map, 117 binary
organ/tissue masks, and a CSV report of dosimetric quantities extracted
using those masks. The reports include absorbed doses for each segmented
organ/tissue, integrated from scan time to infinity and from injection
time to infinity, along with spatial dose standard deviations, dose
conversion factors, scan-time dose rates, target volumes, estimated
masses and voxel counts. Each folder also contains dosimetry config.json,
recording requested simulation histories, completed histories where
available, the integration interval, cumulated activity and available
injection-correction parameters, and metadata.json, recording
radionuclide identity and physical half-life, administered activity,
body weight, acquisition and injection times, PET voxel units and
decay-correction information. Absorbed doses and scan-time dose rates for
the five anatomical regions with the highest median dose in each
radiopharmaceutical group are summarized in
\textbf{Table~\ref{tab:topdose}}. \textbf{Figure~\ref{fig:overlays}} shows
example CT-dose overlays.

The resulting maps represent dose integrated from scan time to infinity,
assuming physical decay alone and an unchanged spatial activity
distribution. Voxel values encode absorbed dose in Gy. Using the recorded
physical half-life and cumulated activity, these maps can be converted to
scan-time dose-rate maps or dose conversion factor maps in Gy per decay.
The latter describe dose per decay across the complete imaged source
distribution. Documented Python tools are provided for selective
downloading and map conversion; conversion outputs include JSON sidecars
recording units, time reference and conversion factors. Detailed
conversion equations are provided in the Supplementary Material.

\section*{Discussion}

DosePET provides 2,100 patient-specific, three-dimensional radiation dose
distributions derived from PET/CT examinations acquired at five
institutions using seven scanner models. This resource could support the
development and evaluation of radiation dose prediction models and
fine-tuning of foundation models for nuclear medicine.

The source anatomical and activity images are publicly available,
allowing users to pair them with the MC dose maps, extract predictors
such as radiomic features and develop predictive models. Agentic systems
have demonstrated automation of medical AI workflows
\citep{ref12,ref13}. DosePET could provide reference data for similar
research workflows targeting rapid organ/tissue dosimetry.

DosePET is publicly available at:
\url{https://huggingface.co/datasets/eltzanis/DosePET}. Python scripts
support selective downloading and dose-map conversion. A companion
SKILL.md guides agents through downloading, matching source PET/CT
examinations and preparing data for modelling.

Users can run the supplied code to convert the NIfTI dose images into
scan-time dose-rate or dose-per-decay maps to suit their research needs.
Moreover, they can apply either the provided AI-generated masks or
manually delineated contours to extract dosimetric quantities from the
original or converted dose maps.

\section*{Acknowledgements}
None.

\section*{Funding}
None

\clearpage
\onecolumn
\begin{figure}[p]
\centering
\includegraphics[width=\linewidth,height=0.78\textheight,keepaspectratio]{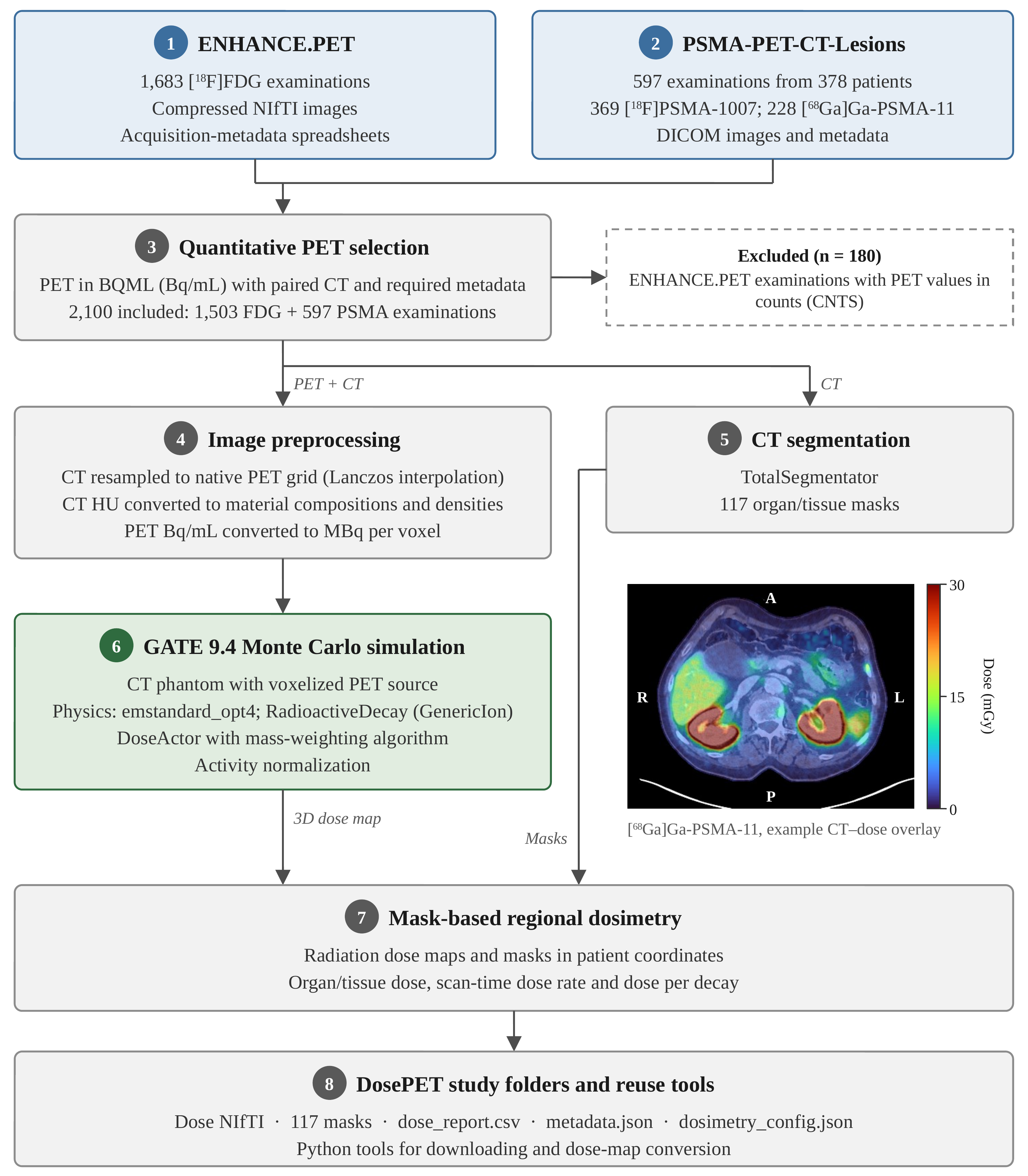}
\caption{DosePET generation workflow. Quantitative PET/CT examinations were selected from ENHANCE.PET and PSMA-PET-CT-Lesions. The CT was resampled to the PET grid to define the voxelized phantom, and PET concentrations were converted to voxel activities for the source. GATE generated the dose field. TotalSegmentator used the original CT to provide anatomical masks for regional extraction. Study folders contain dose maps, anatomical masks, dose reports and metadata. Python tools support downloading and map conversion.}
\label{fig:workflow}
\end{figure}

\clearpage
\section*{Tables}

\begin{table}[h!]
\centering
\caption{Participant and examination characteristics at PET/CT examination.}
\label{tab:characteristics}
\small
\begin{tabular}{@{}>{\raggedright\arraybackslash}p{0.34\linewidth}ccc@{}}
\toprule
\textbf{Characteristic} & \textbf{\FDG{} (ENHANCE.PET)} & \textbf{\FPSMA{}} & \textbf{\GaPSMA{}} \\
\midrule
Number of examinations & 1,503 & 369 & 228 \\
Number of unique participants & N/A & 253 & 166 \\
Age (years) & 60.7 ± 15.0 & 71.7 ± 8.5 & 71.0 ± 7.6 \\
Age range (years) & 11-90 & 48-90 & 50-90 \\
Number of male patients (\%) & 910 (60.5) & 369 (100.0) & 228 (100.0) \\
Number of female patients (\%) & 587 (39.1) & 0 (0.0) & 0 (0.0) \\
Body weight (kg) & 78.6 ± 18.2 & 82.9 ± 14.7 & 82.6 ± 13.2 \\
Administered activity (MBq) & 311.5 ± 64.9 & 246.1 ± 27.1 & 214.5 ± 44.6 \\
\bottomrule
\end{tabular}

\vspace{4pt}
\begin{minipage}{\linewidth}
\footnotesize
Continuous variables are mean ± sample standard deviation, except age range. Age was available for 1,498 FDG examinations and all PSMA examinations. Sex was not reported for six FDG examinations. Repeated examinations contribute separately to examination-level summaries. The PSMA cohort contains 378 unique participants; 41 underwent examinations with both radionuclides, so the subgroup participant counts are not additive. In the PSMA-PET-CT-Lesions dataset, ages above 89 years are encoded as 90 for de-identification; a recorded age of 90 therefore represents 90 years or older.
\end{minipage}
\end{table}

\vspace{1.5em}

\begin{table}[h!]
\centering
\caption{Five anatomical regions with the highest median absorbed radiation dose in each radiopharmaceutical group.}
\label{tab:topdose}
\small
\begin{tabular}{@{}llrcc@{}}
\toprule
\textbf{Tracer} & \textbf{Anatomical region} & \textbf{n} & \textbf{Mean regional dose (mGy)} & \textbf{Scan-time dose rate (mGy/h)} \\
\midrule
\FDG{} & Urinary bladder & 1,503 & 15.22 (10.26–23.39) & 5.77 (3.89–8.86) \\
\FDG{} & Brain & 1,060 & 10.77 (8.76–12.76) & 4.08 (3.32–4.84) \\
\FDG{} & Trachea & 1,503 & 9.85 (8.42–11.65) & 3.73 (3.19–4.42) \\
\FDG{} & Right middle lung lobe & 1,503 & 5.91 (4.33–8.39) & 2.24 (1.64–3.18) \\
\FDG{} & Left upper lung lobe & 1,503 & 5.10 (3.87–7.50) & 1.93 (1.47–2.84) \\
\midrule
\FPSMA{} & Gallbladder & 340 & 13.58 (7.01–24.84) & 5.14 (2.65–9.41) \\
\FPSMA{} & Right kidney & 367 & 13.32 (10.83–17.11) & 5.05 (4.10–6.48) \\
\FPSMA{} & Liver & 369 & 12.18 (9.86–14.65) & 4.61 (3.73–5.55) \\
\FPSMA{} & Left kidney & 367 & 11.98 (9.37–15.37) & 4.54 (3.55–5.82) \\
\FPSMA{} & Spleen & 369 & 9.42 (7.60–11.94) & 3.57 (2.88–4.52) \\
\midrule
\GaPSMA{} & Right kidney & 228 & 23.03 (15.75–30.16) & 14.13 (9.67–18.50) \\
\GaPSMA{} & Left kidney & 227 & 21.80 (15.45–29.90) & 13.37 (9.49–18.36) \\
\GaPSMA{} & Urinary bladder & 228 & 7.42 (4.62–11.43) & 4.56 (2.84–7.02) \\
\GaPSMA{} & Duodenum & 228 & 6.32 (4.72–8.48) & 3.88 (2.90–5.21) \\
\GaPSMA{} & Spleen & 228 & 5.90 (4.51–8.41) & 3.62 (2.78–5.17) \\
\bottomrule
\end{tabular}

\vspace{4pt}
\begin{minipage}{\linewidth}
\footnotesize
Values are median (25th–75th percentile) across examinations. Each examination contributes its mean voxel dose within the indicated mask. Regions are ranked separately within each group by the cohort median of these examination-level means; the top five differ across groups. Empty masks were excluded, and nonempty masks with zero mean dose were retained. Absorbed dose is scan-to-infinity under physical decay alone; rates are at scan start.
\end{minipage}
\end{table}

% ---------------------------------------------------------------------------
% Supplementary Material
% ---------------------------------------------------------------------------
\clearpage
\section*{Supplementary Material -- DosePET dataset}
\addcontentsline{toc}{section}{Supplementary Material}

\subsubsection*{Patient specific Monte Carlo radiation dose calculation}
This supplement describes how the quantitative PET volume and CT-derived
phantom define the simulated radiation field, and how the resulting
scores are converted to scan-referenced dose maps and regional summaries.
The calculations use a single PET time point.

\subsubsection*{1. Image preparation and activity calibration}
The CT was sampled at the physical positions of the native PET voxels
using an identity transform and Lanczos windowed-sinc interpolation.
Voxels outside the CT extent were assigned $-1024$~HU. Image origins,
directions and voxel spacings defined the physical correspondence. The
activity image retained the native PET resolution.

For PET voxel $i$, let $c_{i}$ be its quantitative activity concentration
in Bq/mL. With voxel dimensions $\Delta x$, $\Delta y$ and $\Delta z$ in
millimetres, the voxel volume and assigned activity were
\begin{equation*}
v = \frac{\Delta x\,\Delta y\,\Delta z}{1000}\ \text{mL};\qquad
a_{i} = 10^{-6}\, v \max(c_{i}, 0)\ \text{MBq}
\tag{S1}
\end{equation*}
The activity represented by the complete image was
$A_{0} = 10^{6} \sum_{i} a_{i}$ Bq. The scan-referenced source
distribution was defined by weights $p_{i} = a_{i}/\sum_{i} a_{i}$. Thus,
the simulation used the activity in the imaged volume, rather than
substituting the administered activity. Each history's source voxel was
sampled according to these weights.

For the ENHANCE.PET NIfTI pathway, reconstructed image intensities were
used. Activity calibration and the integration model require a consistent
decay reference. The calculations treated the calibrated PET image as the
activity distribution at scan time. The radionuclide, physical half-life
and relevant acquisition metadata were retained with each examination.

The resampled CT and activity volume underwent the same RAI orientation
conversion and were written as double-precision MetaImage files,
CTRS.mhd/.raw and ACTM.mhd/.raw. Both used a common virtual simulation
frame. Original image geometry was retained separately for conversion of
the scored dose to patient coordinates.
(\href{https://opengate.readthedocs.io/en/v9.4/voxelized_source_and_phantom.html}{GATE voxelized geometry and source documentation}\footnote{\url{https://opengate.readthedocs.io/en/v9.4/voxelized_source_and_phantom.html}})

\subsubsection*{2. Phantom composition and transport configuration}
CT intensities were clipped to [$-10{,}000$, $10{,}000$] HU. GATE's
HounsfieldMaterialGenerator used the supplied density and
elemental-composition tables to generate HUmaterials.db and HU2Mat.txt,
with a density tolerance of 0.005~g/cm\textsuperscript{3}. The phantom
was implemented as an ImageNestedParametrisedVolume. The material table
defined air, lung, adipose tissue, soft tissue and bone compositions.

GATE version 9.4 was run through its Docker image. The voxelized source
used the linear image translator with a scale of 1 MBq. An ion with zero
initial kinetic and excitation energies represented the radionuclide:
$Z = 9$, $A = 18$ for F-18, or $Z = 31$, $A = 68$ for Ga-68. The source
specified an isotropic angular distribution, the forced-unstable flag and
the examination's physical half-life. The calculations used the half-life
recorded for each examination. Decay emissions were generated by the
radioactive-decay model; the zero-energy setting describes the parent
ion.
(\href{https://opengate.readthedocs.io/en/v9.4/source_and_particle_management.html}{GATE source documentation}\footnote{\url{https://opengate.readthedocs.io/en/v9.4/source_and_particle_management.html}})

The macros selected emstandard\_opt4 and added RadioactiveDecay for
GenericIon, together with Decay. Explicit photon and electron
production-range cuts were 0.1 mm in GlobalBox and 1.0 mm in the world.
These cuts are range parameters, not uniform kinetic-energy thresholds.
The electromagnetic table settings specified 0.1~keV–10~GeV, with 220
bins for the stopping-power and interaction-length tables, and the
electron multiple-scattering geometrical step limiter was
distanceToBoundary.

DoseActor was attached to the phantom, with resolution and voxel
dimensions set from the activity image, stepHitType random, and dose
algorithm MassWeighting. Energy deposition was scored in MeV and absorbed
dose in Gy. The mass used for voxel dose scoring was derived from the
simulation phantom.
(\href{https://opengate.readthedocs.io/en/v9.4/tools_to_interact_with_the_simulation_actors.html#dose-measurement-doseactor}{GATE DoseActor documentation}\footnote{\doseactorurl})

\subsubsection*{3. Monte Carlo normalization and temporal model}
Let $b_{j}$ denote the accumulated GATE dose score in target voxel $j$
over $N$ simulated histories. For a scoring voxel with mass $m_{j}$, this
represents deposited energy divided by mass, with the energy expressed in
joules and mass in kilograms. It is an accumulated score, not a dose
already normalized to the patient's activity.

Parallel workers used MersenneTwister with explicit seeds
$k = 1,\ldots,K$. Each worker was assigned the integer part of
$N_{\text{requested}}/K$ histories. Additive dose and energy-deposition
arrays were summed; voxel masses were retained once because the phantom
mass does not increase with the number of histories. The normalization
used the completed-history count recorded by the pipeline. The count must
correspond to the scores included in the merged field. The requested
count was $10^{7}$ histories per examination; the completed count may
differ because of worker allocation.

For the examination's physical half-life $T_{1/2}$, define
$\lambda = \ln(2)/T_{1/2}$. With $\tau$ measured from scan acquisition,
the model assumed that every source voxel decreased by the same
physical-decay factor. The integrated number of decays represented by the
image was therefore
\begin{equation*}
\tilde{A} = \int_{0}^{\infty} A_{0}\, e^{-\lambda\tau}\, d\tau = \frac{A_{0}}{\lambda}\ \text{Bq s}
\tag{S2}
\end{equation*}
The dose and scan-time dose rate at voxel $j$ follow from the mean score
per simulated decay:
\begin{equation*}
D_{j} = \frac{b_{j}}{N}\,\tilde{A}\ \text{Gy};\qquad
\dot{D}_{j}(0) = \frac{b_{j}}{N}\,A_{0} = \lambda D_{j}\ \text{Gy/s}
\tag{S3}
\end{equation*}
This separates transport from temporal integration. Multiplication by
$A_{0}$ gives the dose rate for the measured source distribution, whereas
multiplication by $A_{0}/\lambda$ gives its physical-decay-only total
radiation dose. The latter assumes unchanged anatomy and relative
source-voxel weights after acquisition. No biological clearance or
redistribution is fitted from these single-time-point images. Activity
outside the imaged field is not represented, and the maps do not include
dose from the CT acquisition.

The code also supports a timing-based optional extrapolation toward
injection. When its timing and decay-reference conditions are met, an
additional factor $\exp(\lambda\Delta t)$ is applied, where $\Delta t$ is
the injection-to-scan interval. Those optional values are model
extrapolations and do not measure the earlier uptake history.

\subsubsection*{4. Dose export and target summaries}
Organ and tissue masks were generated automatically using
TotalSegmentator with the total CT task. The supplied batch invoked GPU
inference with fastest=False and preferred the original CT volume as
input. The generated binary masks provided the target regions for dose
extraction; the segmentation did not define or restrict the PET emission
source.
(\href{https://doi.org/10.1148/ryai.230024}{TotalSegmentator software paper}\footnote{\url{https://doi.org/10.1148/ryai.230024}})

The dose field must be placed in patient coordinates before regional
extraction. In the supplied standard export pathway, dose was linearly
resampled to the reference CT grid and stored as a compressed NIfTI image
with integer storage and NIfTI scaling. Physical dose values require
applying that scaling. Binary target masks use nearest-neighbor
interpolation when sampling onto another grid. Grid compatibility
includes origin, direction, spacing and dimensions.

For target $R$ containing $n_{R}$ equal-volume mask voxels, the reported
mean is the arithmetic mean of sampled voxel doses:
\begin{equation*}
\bar{D}_{R} = \frac{1}{n_{R}} \sum_{j \in R} D_{j};\qquad
V_{R} = n_{R}\, v_{R}
\tag{S4}
\end{equation*}
Here $v_{R}$ is the extraction-grid voxel volume. The dose includes
contributions from the full PET source map, including activity in the
target and elsewhere. Minimum, maximum and spatial dose standard
deviation were also calculated within each mask.

The target-dose coefficient and mean scan-time dose rate were calculated
as
\begin{equation*}
C_{R} = \frac{\bar{D}_{R}}{\tilde{A}}\ \text{Gy/(Bq s)};\qquad
\bar{\dot{D}}_{R}(0) = \lambda\, \bar{D}_{R}\ \text{Gy/s}
\tag{S5}
\end{equation*}
The coefficient is conditional on the complete normalized source
distribution and patient anatomy; it is not a source-organ-specific
S-value. CSV coefficients in Gy/(Bq s) are multiplied by $10^{15}$ when
expressed as fGy/(Bq s) in the aggregate table. Dose rates in Gy/s are
multiplied by $10^{3}$ for mGy/s and by $3.6 \times 10^{6}$ for mGy/h.

\subsubsection*{5. Statistical uncertainty}
The following expressions define uncertainty calculations conditional on
retained history scores. For a voxel's individual history contributions
$s_{hj}$, the relative standard error of the mean score can be written
\begin{equation*}
u_{j}^{2} = \frac{\sum_{h=1}^{N} \left(s_{hj} - \bar{s}_{j}\right)^{2}}{N(N-1)\,\bar{s}_{j}^{2}};\qquad
\bar{s}_{j} = \frac{1}{N}\sum_{h=1}^{N} s_{hj}
\tag{S6}
\end{equation*}
The expression requires $N > 1$ and a nonzero mean.
(\href{https://opengate.readthedocs.io/en/v9.4/tools_to_interact_with_the_simulation_actors.html#dose-measurement-doseactor}{GATE uncertainty documentation}\footnote{\doseactorurl})

\subsubsection*{6. Released map-conversion utilities}
The DosePET repository includes download\_dataset.py for selecting and
downloading studies from the archive manifest, and
tools/convert\_dose\_images.py with a detailed README for working with the
downloaded maps. The converter resolves $T_{1/2}$ from
dosimetry\_config.json or metadata.json and obtains the scan-to-infinity
cumulated activity from cumulated\_from\_scan in the dosimetry
configuration.

For voxel dose $D_{j}$, scan-start rate is $\lambda \cdot D_{j}$ and the
voxel dose coefficient is $D_{j}/\tilde{A}$. The rate command supports
Gy/s, mGy/s, mGy/min and mGy/h. The dcf command outputs Gy per decay of
the complete imaged activity distribution.

The default header mode preserves stored integer voxel values and
rescales the NIfTI slope and intercept. A float32 mode is also provided.
Each converted map has a JSON sidecar recording its quantity, units, time
point, scale factor and the origins of the constants used. Converted
files are written to a separate output location. Original dose-map
geometry is preserved by scalar conversion.

\end{document}